\def\DI{{D_{\text{I}}}}
\def\Dh{{D_{\text{H}}}}
\def\sSW{{\sigma_{i_{\text{SW}}}}}
\def\iSW{{i_{\text{SW}}}}

\documentclass[%
 reprint,
 superscriptaddress,
 preprintnumbers,
 amsmath,amssymb,
 aip,
]{revtex4-1}

\usepackage[utf8]{inputenc}
\usepackage{graphicx}
\usepackage{grffile}
\usepackage[usenames,dvipsnames]{xcolor}
\usepackage{amsmath}
\usepackage[normalem]{ulem}
\usepackage[resetlabels, labeled]{multibib}
\usepackage{appendix}
\newcites{S}{References Supplementary Materials}
\definecolor{orange}{rgb}{1,0.5,0}
\definecolor{goodgreen}{rgb}{0.1,0.5,0}
\definecolor{goodred}{rgb}{0.7,0,0}
\usepackage{lineno}
\setpagewiselinenumbers
\modulolinenumbers[5]

\usepackage[colorlinks,urlcolor=goodgreen,citecolor=blue,linkcolor=goodred]{hyperref}
\usepackage{float}
\usepackage{babel}
\graphicspath{ {images/} }

\makeatother

\begin{document}


\title[SCD in anomalous JJs]{Switching current distributions in ferromagnetic anomalous Josephson junctions}

\newcommand{\orcid}[1]{\href{https://orcid.org/#1}{\includegraphics[width=8pt]{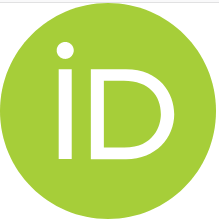}}}
\author{C. Guarcello\orcid{0000-0002-3683-2509}}
\email{Author to whom correspondence should be addressed: cguarcello@unisa.it}
\affiliation{Dipartimento di Fisica ``E.R. Caianiello'', Universit\`a di Salerno, Via Giovanni Paolo II, 132, I-84084 Fisciano (SA), Italy}
\affiliation{INFN, Sezione di Napoli Gruppo Collegato di Salerno, Complesso Universitario di Monte S. Angelo, I-80126 Napoli, Italy}
\author{F.S. Bergeret\orcid{0000-0001-6007-4878}}
\email{fs.bergeret@csic.es }
\affiliation{Centro de Física de Materiales, Centro Mixto CSIC-UPV/EHU, Paseo Manuel de Lardizabal 5, 20018 San Sebastián, Spain}
\affiliation{Donostia International Physics Center, Paseo Manuel de Lardizabal 4, 20018 San Sebastián, Spain}%
\author{R. Citro\orcid{0000-0002-3896-4759}}
\email{rocitro@unisa.it}
\affiliation{Dipartimento di Fisica ``E.R. Caianiello'', Universit\`a di Salerno, Via Giovanni Paolo II, 132, I-84084 Fisciano (SA), Italy}
\affiliation{INFN, Sezione di Napoli Gruppo Collegato di Salerno, Complesso Universitario di Monte S. Angelo, I-80126 Napoli, Italy}
\affiliation{CNR-SPIN c/o Universit\'a degli Studi di Salerno, I-84084 Fisciano (Sa), Italy}


\begin{abstract}

We investigate the switching current distributions of ferromagnetic anomalous Josephson junctions subjected to a linearly increasing bias current. Our study uncovers a significant correlation between the position of the switching current distributions and crucial system parameters, such as the strength of the spin-orbit coupling and the Gilbert damping parameter. This indicates that these parameters can be directly determined through experimental measurements. By conducting a comprehensive analysis of the interplay among noise, magnetization, phase dynamics, and the statistical properties of the switching current distribution, we deepen our understanding of these intriguing cryogenic spintronics devices. These findings hold potential for applications in the field of quantum computing architectures and information processing technologies.
\end{abstract}

\maketitle

Ferromagnetic anomalous Josephson junctions (FAJJs), formed by combining superconductivity with ferromagnetic materials, have garnered significant attention due to their unique properties and potential applications in spintronics, quantum technologies, and information processing~\cite{Feo10,Rom13,Lin15,Esc15,Gin16,Gol17}. These junctions exhibit intriguing phenomena, including anomalous phase shifts and magnetization reversal induced by the flow of supercurrent, making them promising candidates for a wide range of cutting-edge technological advancements.

The combination of superconductivity and ferromagnetism in these junctions offers exciting possibilities for spintronics applications~\cite{Mel22}. The exchange interaction between the superconducting and ferromagnetic regions enables the control of the magnetization direction through the manipulation of superconducting phases. This unique functionality holds great promise for the development of spintronics memory elements, magnetic field sensors, and logic circuits with low power consumption and high-speed operation.

Since the work by Shukrinov \emph{et al.}~\cite{Shu17}, several studies have explored the dynamics of FAJJs, for instance focusing on the current-induced magnetization reversal phenomenon, in both short junctions~\cite{Nas18,Shu18b,Shu19,Maz20,Shu20,Bel22,Jan22,Abd22,Nas22,Bot23,Sam23} and superconducting quantum interference devices (SQUIDs)~\cite{Shu18,Mir20,GuaCit20}. An interested reader may refer to Refs.~\onlinecite{Shu22,Bob22} for a comprehensive review and further details.

\begin{figure}[b!!]
\includegraphics[width=\columnwidth]{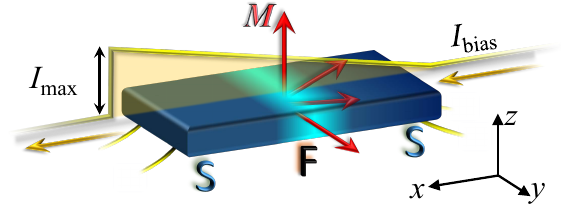}
\caption{Sketch of the AFJJ driven by a linearly ramping bias current, $I_{\text{bias}}(t)$.}\label{Figure01}
\end{figure}

Noise influences the behavior of these systems, playing an important role in both understanding their operation and harnessing their full potential, although its effect has been little studied to date. Nonetheless, stochastic thermal fluctuations can profoundly affect the behavior and the performance of these devices~\cite{Gua20,GuaBer21}. Understanding the interplay between noise and the complex dynamics of AFJJs is essential for optimizing their operation and ensuring the reliability of spintronics and quantum devices. \\
In the broader context of Josephson junctions (JJs), the study of the switching current distribution (SCD) is of fundamental importance to characterize their behaviour~\cite{Bla16}. It provides insights into the dynamics of the superconducting phase and the interplay between noise and junction parameters, and it was throughly discussed, both theoretically and experimentally, in many works about Josephson devices~\cite{Wallraff2003,Blackburn2014,Bla16,Massarotti2019,Revin2021}. The SCD in conventional JJs can exhibit a wide range of behaviors, ranging from narrow and well-defined distributions to broad and multimodal distributions. In fact, noise sources introduce fluctuations in the superconducting order parameter, leading to variations in the switching process. Consequently, the shape of the SCD reflects the intertwining of noise effects and the specific characteristics of the junction.

Characterizing the SCD and understanding the role of noise in shaping its properties are crucial for designing and optimizing the performance of FAJJs for their various applications. In this Letter, we analyze the impact of thermal noise on the behavior of FAJJs, with a focus on the SCDs. In particular, we demonstrate how they can reveal information on intrinsic characteristics of the junction, such as the Rashba coupling strength and the phenomenological Gilbert damping parameter.
This work adds a fundamental piece to the study of noise-induced effects in magnetic Josephson systems, which plays a key role in understanding the actual response of a real device; indeed, we focus on the switching dynamics under a linearly ramping bias current, moving significantly away from the conventional scenarios considered so far, in which the phenomenon of pulse-controlled magnetization reversal~\cite{Pankratov08,Shu17,Maz20,Gua20,GuaBer21,Bel22,Sam23} and the response to ac currents~\cite{Nas18,Shu19,Abd22} had been largely considered.

The setup that we consider consists of a FAJJ, see Fig.~\ref{Figure01}, with a thin ferromagnetic film with an out-of-plane magnetic anisotropy and a Rashba-like SOC~\cite{Gua20}. 

Due to the interplay between the exchange field and the SOC, the ferromagnetic junction presents a current-phase relation with the form\cite{Buz08,Ber15} $I_{\varphi}=I_c\sin(\varphi-\varphi_0)$. Here, $I_c$ is the critical current of the junction, $\varphi$ is the Josephson phase difference, and $\varphi_0$ is the \emph{anomalous phase shift}, which has been experimentally measured first in Josephson devices including topological insulators and Al/InAs heterostructures or nanowires~\cite{Szo16,Ass19,May19,Str20}.
The value of $\varphi_0$ depends on different system parameters, such as the Rashba coupling $\alpha$~\cite{Ras60,Byc84}, the transparency of S/F interfaces, the spin relaxation, and the disorder degree. For our purposes, the exact dependence of $\varphi_0$ on these parameters is not as relevant as the device geometry. If we assume a two-dimensional SOC with magnetic momenta in the plane of the F film (in particular, we are assuming a monodomain ferromagnetic arrangement), and the bias current flows in $x$-direction, the phase shift $\varphi_0$ is proportional to the $y$-component of the magnetic moment according to~\cite{Buz08,Kon09,Kon15,Ber15}
\begin{equation}\label{phi0}
\varphi_0=r \,m_y,
\end{equation}
where $m_y=M_y/M$, with $M=\sqrt{M_x^2+M_y^2+M_z^2}$ being the modulus of the magnetic moment, and the parameter $r$ depends on $\alpha$ (for the explicit dependence of $r$ on various system parameters, including the strength of the spin-orbit interaction, refer to Refs.~\onlinecite{Buz08,Kon09}). 
Equation~\eqref{phi0} evidently establishes a direct coupling between the magnetic moment and the supercurrent.

The time evolution of the magnetic moment can be described in terms of the Landau–Lifshitz–Gilbert (LLG) equation~\cite{Lan35,Gil04}
\begin{equation}\label{LLG}
\frac{d\textbf{M}}{d\tau}=\frac{\gamma}{M}\left ( \textbf{M}\times\frac{d\textbf{M}}{d\tau} \right )-g_r\textbf{M}\times\textbf{H}_{\text{eff}},
\end{equation}
where $g_r$ denotes the gyromagnetic ratio. The first term on the right-hand side accounts for the dissipation through the phenomenological dimensionless Gilbert damping parameter $\gamma$, while the second term represents the precession around $\textbf{H}_{\text{eff}}$, which components can be calculated as~\cite{Lif90}
\begin{equation}\label{EffectiveField_FreeEnergy}
H_{\text{eff},i}=-\frac{1}{V}\frac{\partial \mathcal{F}}{\partial M_i}, \qquad\text{with}\quad i=x,y,z.
\end{equation}
Here, $V$ is the volume of the F layer and
\begin{equation}\label{FreeEnergy}
\mathcal{F}=-E_J\varphi I_{bias}+E_s(\varphi,\varphi_0)+E_M
\end{equation}
is the free energy of the system.
Here, $E_J=\Phi_0I_c/(2\pi)$ (with $\Phi_0$ being the flux quantum), $I_{bias}$ is the external current in units of $I_c$, $E_s(\varphi,\varphi_0)=E_J[1-\cos(\varphi-\varphi_0)]$, and $E_M=-\frac{\mathcal{K}V}{2}\left ( \frac{M_z}{M} \right )^2$ is the magnetic energy that depends on the anisotropy constant $\mathcal{K}$. 
In the following, we use the parameter $\varepsilon =E_J/(\mathcal{K}V)$ to indicate the ratio between the energy scales of the system. 
In Eq.~\eqref{LLG} we are neglecting a second-derivative term proportional to an angular momentum relaxation time $\tau_a$~\cite{Ciornei2011,Thonig2017,Cherkasskii2020,Neeraj2021}. On timescales shorter than $\tau_a$, this term is demonstrated to give nutation oscillations on top of the precession motion. Instead, on timescales longer than $\tau_a$, the usual LLG equation is expected to work well in describing the magnetic evolution. This is exactly our case, being $\tau_a$ well below both the inverse Josephson frequency and the ramp time, $t_{\text{max}}$

\begin{figure*}
\includegraphics[width=2\columnwidth]{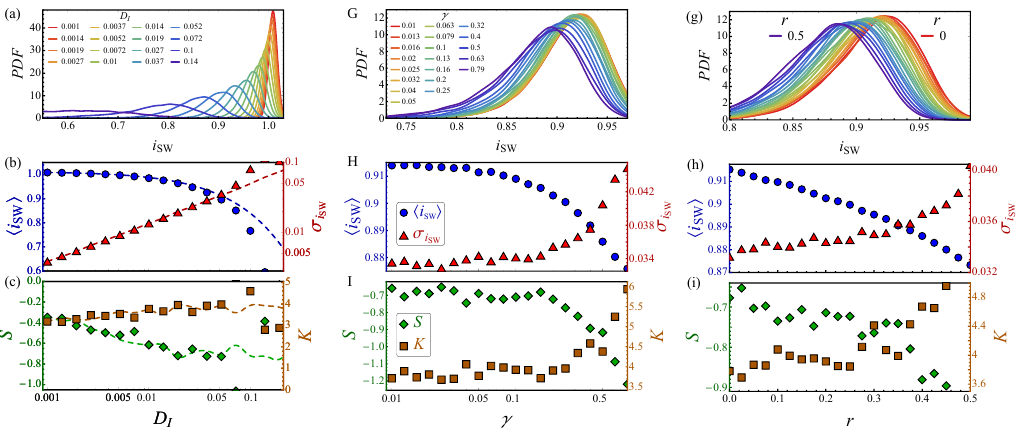}
\caption{PDF of the switching currents $\iSW$, see top panels, moments of the SCDs, i.e., mean values ($\left< \iSW \right>$) and root mean square ($\sSW$) in middle panels, whereas skewness ($S$) and kurtosis ($K$) in bottom panels, as a function of: (a-c) $D_I$ with $(r,\gamma)=(0.25,0.1)$, (d-f) $\gamma$ with $(D_I,r)=(0.05,0.1)$, and (i-l) $r$ (from $r=0$, red curve, to $r=0.5$, purple curve, with $\Delta r=0.02$) with $(D_I,\gamma)=(0.05,0.1)$. The dashed lines in panels (b-c) indicate the SCD moments obtained in the decoupled case, i.e., $r=0$. The legend in panel (e) [(f)] refers also to (b) and (h) [(c) and (i)].}\label{Figure02}
\end{figure*}

From Eq.~\eqref{EffectiveField_FreeEnergy}, we obtain the effective magnetic field
\begin{equation}\label{effectivefield}
\textbf{H}_{\text{eff}}=\frac{\mathcal{K}}{M}\left [ \varepsilon r \sin\left ( \varphi-rm_y \right )\hat{y}+m_z\hat{z} \right ],
\end{equation}
where $m_{x,y,z}=M_{x,y,z}/M$ are the normalized components of the magnetic moment that have to satisfy the condition $m_x^2+m_y^2+m_z^2=1$. The LLG equations can be conveniently expressed in spherical coordinates~\cite{Rom14}, so as to conserve the magnetic moment modulus even when stochastic noise contributions are taken into account. Then, $m_{x,y,z}$ can be written in terms of polar and azimuthal angles $\theta$ and $\phi$ as
\begin{eqnarray}\label{Msphericalcoord}\nonumber
m_x(\tau)&=&\sin\theta(\tau)\cos\phi(\tau)\\
m_y(\tau)&=&\sin\theta(\tau)\sin\phi(\tau)\\\nonumber
m_z(\tau)&=&\cos\theta(\tau).
\end{eqnarray}
We can also define the $\theta$ and $\phi$ components of the normalized effective field as
\begin{eqnarray}\label{AngleFieldsEffective-a}
&&\widetilde{H}_{\text{eff},\theta}=\varepsilon r\sin(\varphi-rm_y)\cos\theta\sin\phi-m_z\sin\theta\qquad\\\label{AngleFieldsEffective-b}
&&\widetilde{H}_{\text{eff},\phi}=\varepsilon r\sin(\varphi-rm_y)\cos\phi.
\end{eqnarray}
Thus, if the time is normalized to the inverse of the ferromagnetic resonance frequency $\omega_F=g_r \mathcal{K}/M$, that is $t=\omega_F\tau$, the LLG equations in spherical coordinates reduce to the following two coupled equations~\cite{Rom14}
\begin{eqnarray}\label{LLGsphericalcoord_a}
\frac{\mathrm{d} \theta}{\mathrm{d}t}=&&\frac{1}{1+\gamma^2}\left ( \widetilde{H}_{\text{eff},\phi}+\gamma\; \widetilde{H}_{\text{eff},\theta}\right )\\\label{LLGsphericalcoord_b}
\sin \theta \frac{\mathrm{d} \phi}{\mathrm{d} t}=&&\frac{1}{1+\gamma^2}\left (\gamma\; \widetilde{H}_{\text{eff},\phi}- \widetilde{H}_{\text{eff},\theta} \right ).
\end{eqnarray}

The dynamics of the Josephson phase can be described in terms of the resistively and capacitively shunted junction (RCSJ) model~\cite{Bar82,Gua19,Gua20PRR,Gua21}, which can be generalized to include the anomalous phase shift $\varphi_0=rm_y$~\cite{Rab19}:
\begin{subequations}\label{RSJ}\begin{align}
\frac{d\varphi}{d t}&=&\!\!\!\!\omega\chi+r\frac{d m_y }{d t}\qquad\qquad\qquad\qquad\qquad\qquad\quad\\
\frac{d\chi }{d t}&=&\!\!\!\!\frac{\omega}{\beta_c} \left [ I_{bias}(t)+I_{\text{th}}(t)-\sin\left ( \varphi-rm_y \right ) -\chi \right ].
\end{align}
\end{subequations}
Here, time ($t$) is still in units of the inverse of the ferromagnetic resonance frequency, and $\omega=\omega_c/\omega_F$, with $\omega_c=2\pi I_c R/\Phi_0$ being the characteristic frequency~\cite{Bar82} of the junction with a normal-state resistance $R$. The damping is quantified by the McCumber parameter, $\beta_c=2\pi I_cCR^2/\Phi_0$ ($C$ is the junction capacitance), which is related to the so-called \emph{quality factor} of the junction according to $\beta_c=Q^2$. In this work, we consider a moderately underdamped junction, i.e., we set $Q=5$ just in line with Refs.~\onlinecite{Shu20,Jan22,Bot23}. The values of the other parameters are $\varepsilon =10$ and $\omega=1$. The value of energy ratio $\varepsilon$ is expected to range from $\sim100$, for a weak magnetic anisotropy, to $\sim1$, for a strong anisotropy~\cite{Shu17}. We chose an intermediate value, in line with other theoretical work~\cite{Shu17}, but to guess how a different value affects the system we observe that  $\varepsilon$ comes into play in \emph{i}) the normalised effective field, see Eqs.~(\ref{AngleFieldsEffective-a}-\ref{AngleFieldsEffective-b}) and in \emph{ii}) the thermal field intensity. Thus, we expect a larger (smaller) $\varepsilon$ to give faster (slower) magnetic dynamics and a more (less) relevant noise-induced effect, e.g., see Refs.~\onlinecite{Shu17,Shu18b,Maz20,Shu22,Bob22}.

The noise term $I_{\text{th}}(t)$ is a sort of ``thermal current'' with the usual white-noise statistical properties that, in normalized units, can be expressed as~\cite{Bar82,Gua13,Gua16}
\begin{equation}
\left < I_{\text{th}}(t) \right >=0\qquad\qquad
\left < I_{\text{th}}(t)I_{\text{th}}({t}') \right >=2\DI \delta \left ( t-{t}' \right ),
\label{thermalcorrelatorI}
\end{equation}
with 
\begin{equation}\label{thermalcurrentamplitude}
\DI =\frac{k_BT}{R}\frac{\omega_F}{I_c^2}=\frac{1}{\omega}\frac{k_BT}{ E_J},
\end{equation}
being the dimensionless intensity of thermal current fluctuations.
In this work we consider also magnetic field fluctuations, i.e., we take into account in Eq.~\eqref{LLG} a delta-correlated stochastic ``thermal field'' $H_{\text{th}}$ contribution, with an intensity $\Dh =(\gamma\, \varepsilon \omega)\DI$ (see Ref.~\onlinecite{Gua20} for more details). This means that one can regulate the relative strength of the two noisy mechanisms by changing the magnetization energy, the Gilbert damping parameter, or the magnetic-resonance frequency. Thus, the system parameters can be optimized in such a way to make, for instance, the impact of the thermal field negligible with respect to the thermal current. Below, although we only specify the value of $\DI$, we point out that both thermal current and thermal field are included.

In order to construct a SCD, we consider a linearly increasing bias current $i_{bias}(t)=I_{bias}(t)/I_c=v_bt$, where $v_b=t_{\text{max}}^{-1}$ is the ramp speed. A measurement consists in slowly and linearly ramping the bias current in a time $t_{\text{max}}$, so that $I_{\text{max}}\equiv I_b(t_{\text{max}})=I_c$, and to record the current value, $\iSW$, at which a switch occurs, namely, at which the phase particle leaves the initial potential well. Alternatively, one could look at the voltage drop. In this way, it should be also possible to distinguish phase diffusion events, in which the phase particle once escaped is retrapped in an adjacent minimum due to damping affecting its motion. In this case, the resulting SCD should show opposite behavior with temperature, i.e., its width should increase as the temperature decreases~\cite{Revin20,Pankratov22}.
In this readout scheme, the noise influence is considered in the limit of the adiabatic bias regime, where the change of the slope of the potential induced by the bias current is slow enough to keep the phase particle in the metastable state until the noise pushes out the particle. Since we are dealing with stochastic switching processes, we need to perform many independent experiments under the same conditions, in order to obtain comprehensive statistics. In particular, sequences of $N=10^4$ independent numerical experiments of maximum duration $t_{\text{max}}=10^4$ are performed. Thus, the collection of these accumulated switching currents forms a SCD, whose moments, i.e., mean, variance, skewness, and kurtosis, can be investigated. 

We assume that the magnetic moment initially points towards the $z$-direction, that is $\textbf{M}=(0,0,1)$ at $t=0$, and that $\varphi(0)= d\varphi(0)/dt=0$. With these initial conditions, we solve Eqs.~\eqref{LLGsphericalcoord_a}-\eqref{RSJ} self-consistently, at different system parameter values. Once the switching has taken place, the Josephson phase undergoes fast $2\pi$ rotations and a non-zero voltage drop appears. In this condition, the instantaneous value of magnetic moment is not so relevant, but rather the time-averaged value it takes after switching. Computing also the ensemble average on the total amount of independent realizations, we observe, in the regions of parameter space under consideration, zero average after-switching magnetic moments (data not shown). The readout of the magnetic state of the system could be eventually achieved through a similar nondestructive scheme proposed in Ref.~\onlinecite{Gua20}.

In Fig.~\ref{Figure02} we collect the probability distribution function (PDF) of switching currents $\iSW$ (see top panels) and the moments (i.e., mean value, $\left< \iSW \right>$, and root mean square, $\sSW$, in the middle panels, while skewness, $S$, and kurtosis, $K$, are shown in the bottom panels) of the SCDs, varying the main system parameters, $(D_I,r,\gamma)$, in suitable ranges. In particular, we choose: in panels (a-c) $D_I\in[0.001-1.0]$ with $(r,\gamma)=(0.25,0.1)$, in (d-f) $\gamma\in[0.01-1.0]$ with $(D_I,r)=(0.05,0.1)$, and in (i-l) $r\in[0-0.5]$ with $(D_I,\gamma)=(0.05,0.1)$. Skewness and kurtosis are obtained as $$\tilde{S}[X]\!=\!\textup{E}\left [ \left ( \frac{X-\mu}{\sigma} \right )^3 \right ]\;\;\text{and}\;\;\tilde{K}[X]\!=\!\textup{E}\left [ \left ( \frac{X-\mu}{\sigma} \right )^4 \right ]$$ where $X$ represents the random variable of the measurements, $E[\cdot ]$ is the expectation operator, $\mu=\textup{E}\left [ X \right ]$, and $\sigma^2=\text{var}\left ( X \right )$. These quantities can be estimated through the measured switching currents $i_{SW,j}$ as $$S\!=\!\frac{\left < \left ( i_{SW,j}-\left < i_{SW} \right > \right )^3 \right >}{\sigma^3}\;\;\text{and}\;\; K\!=\!\frac{\left < \left ( i_{SW,j}-\left < i_{SW} \right > \right )^4 \right >}{\sigma^4},$$ where $\left < i_{SW} \right >$ and $\sigma$ are the estimates of the mean and the standard deviation of the switching distribution, respectively.

We first look in Fig.~\ref{Figure02}(a) at the behavior of the SCDs at different noise intensities $D_I$, which is proportional to the temperature according to Eq.~\eqref{thermalcurrentamplitude}. We note that the $D_I$ values are chosen in such a way that the moments and the magnetic moments in Figs.~\ref{Figure02}(b-c) appear equally spaced on a logarithmic scale. The more the noise intensity increases, the more the SCDs are lower, wider, and centered at smaller switching current values. The first two moments, i.e., mean value, $\left< \iSW \right>$, and root mean square, $\sSW$, see panel (b), show a typical trend: in particular, $\left< \iSW \right>$ tends to saturate at the level $\sim1$ at low noises and then it decreases, while $\sSW$ increases exponentially (i.e., linearly on a log-log scale) as the noise increases. We stress that we are assuming to work at temperatures above the so-called ``crossover temperature''~\cite{Aff81,Bla16}, i.e., that value below which macroscopic quantum tunneling (MQT) effects dominate the switching dynamics: in this case, being the MQT probability independent of $T$, one would obtain overlapping SCDs, all centered around the crossover temperature. In panel (c) we show the third and fourth moments: we see that the skewness $S$ (the kurtosis $K$) takes the value $S\sim-0.4$ ($K\sim3$) for $D_I=0.001$ and decreases (increases) up to the value $S\sim-0.8$ ($K\sim4$) for $D_I=0.05$. The obtained values are in line with those already observed for other Josephson systems~\cite{Mur13}. In other words, increasing noise makes the SCDs more asymmetric and with tails more pronounced.

In order to understand the interplay between the Josephson and the magnetic systems, we enrich panels (b) and (c) with the SCD moments of a conventional JJ [i.e., setting $r=0$ in Eq.~\eqref{RSJ} in order to decouple the Josephson and magnetic systems] marked with dashed lines. It is clear that all moments tend to diverge from the pure-thermal behavior for noises larger that $\DI\gtrsim0.05$. The response of a FAJJ therefore cannot be merely assimilated to that of a conventional JJ residing at a slightly higher temperature, i.e., subject to a larger noise intensity. Indeed, in principle, one could assume an effective temperature such that the behavior of the SCDs in Fig.~\ref{Figure02}(a-c) could again be achieved in a conventional JJ. In particular, we could search for an ad-hoc additional noise intensity that gives the values of $\left< \iSW \right>(D_I)$ that agree with those we obtained; however, by doing so, the other moments would differ significantly from those presented in Fig.~\ref{Figure02}(b-c) (data not shown). Thus, one cannot merely reduce our results to what would be observed considering a conventional JJ with a higher effective temperature.

The central panels of Fig.~\ref{Figure02} report how the SCDs depend on the Gilbert damping parameter, $\gamma\in[0.01-1.0]$, imposing $(D_I,r)=(0.05,0.1)$. Looking at the PDF of the switching currents in Fig.~\ref{Figure02}(d), it is evident that by changing $\gamma$ both the position and the shape of the SCDs change. Specifically, $\left< \iSW \right>$ ($\sSW$) show a monotonically decreasing (increasing) behavior increasing $\gamma$, see Fig.~\ref{Figure02}(e). Instead, both $S$ and $K$ remain practically unchanged for $\gamma\lesssim0.2$, while for higher $\gamma$ they tend to increase, the former, and to decrease, the latter, see Fig.~\ref{Figure02}(f). The value of $r$ significantly affects the threshold $\gamma$ value above which the shape of the SCDs starts to change appreciably. We stress that the scattered behavior of the higher-moment curves can be smoothed out by increasing the number of numerical repetitions.

Finally, Fig.~\ref{Figure02}(g-h) says that by increasing the SOC strength, $r$, the SCDs tend to shift towards lower switching currents. In particular, we observe that both $\left< \iSW \right>$ and skewness (root-mean-square and kurtosis) tend to decrease (increase) almost linearly by increasing $r$.

Figure~\ref{Figure02} says that the analysis of both the average and the shape of the switching current distributions could be employed to experimentally determine $\gamma$ or $r$, the other being known.

In conclusion, we discussed the behavior of the SCDs of a current-biased $\varphi_0$-junction, that is a superconductor-ferromagnet-superconductor JJ with a Rashba-like spin-orbit coupling. We investigated the noise impact on both the Josephson phase dynamics and the magnetization of the system in the case of a slowly, linearly ramping bias current. 
In particular, we showed how a change in the various characteristic parameters of the system impacts the SCDs, also looking at the distribution moments, i.e., mean values, root mean square, skewness, and kurtosis. In particular, their analysis give information on the main quantities of the system, i.e., the phenomenological Gilbert damping parameter $\gamma$ and the SOC strength. However, the description of a concrete experiment should first proceed with a fine-tuning of the other parameters, such as $I_c$, $\varepsilon$, the saturation magnetization, $\omega_c$, and $\omega_F$ (some values can be found in, e.g., Refs.~\onlinecite{Shu17,Nas18, Shu18b, Nas22, Shu22,Bob22}).

The ability to control and manipulate the magnetization states through the superconducting phases opens up possibilities for encoding and processing quantum information, thus our results may contribute to the development of optimized spintronics.


\begin{acknowledgments}
F.S.B. acknowledges financial support from Spanish MCIN/AEI/ 10.13039/501100011033 through project PID2020-114252GB-I00 (SPIRIT) and TED2021-130292B-C42, and the Basque Government through grant IT-1591-22. R.C. acknowledges the project HORIZON-EIC-2022-PATHFINDERCHALLENGES-01 GA N.101115190–IQARO.
\end{acknowledgments}

\section*{Data Availability Statement}
The data that support the findings of this study are available from the corresponding author upon reasonable request.

\bibliography{biblio}

\end{document}